\documentclass[%
aip,
jap,%
amsmath,amssymb,
reprint,%
]{revtex4-1}

\usepackage{graphicx}
\usepackage{dcolumn}
\usepackage{bm}
\begin{document}

\title{Magnetodielectric and spin-lattice coupling in quasi 1D Ising spin chain CoNb$_{2}$O$_{6}$}

\author{M. Nandi} 
\affiliation{Saha Institute of Nuclear Physics, 1/AF Bidhannagar, Calcutta 700 064, India}
\author{D. Prabhakaran}
\affiliation{Clarendon Laboratory, Department of Physics, University of Oxford, Oxford OX1 3PU, UK}
\author{P. Mandal}
\affiliation{Saha Institute of Nuclear Physics, 1/AF Bidhannagar, Calcutta 700 064, India}
\email{prabhat.mandal@saha.ac.in}
\date{\today}

\begin{abstract}
We have studied magnetodielectric and spin-lattice coupling in CoNb$_{2}$O$_{6}$ single crystals. Magnetostriction and magnetodielectric experiments are performed at temperatures in and above antifferomagnetic phase of quasi 1D Ising spin chain CoNb$_{2}$O$_{6}$. Field induced magnetic transitions are clearly reflected in magnetodielectric measurement as well as magnetostriction measurement also. Two sharp anomalies are found around the critical fields of antiferromagnetic to ferrimagnetic transition and ferrimagnetic to saturated paramagnetic transition in both magnetodielectric and magnetostriction experiments. High field anomaly is more pronounced for magnetodielectric response and magnetostriction also. So, in CoNb$_{2}$O$_{6}$, spins are strongly coupled with lattice as well as charges also. 
\end{abstract}


\maketitle
\section{Introduction}
 In recent years geometrically frustrated triangular lattice systems have attracted immense interest due to its different kind of magnetic phase transitions and degenerate ground states. Several triangular lattice systems also exhibit multiferroic behavior.\cite{seki} Geometrical frustration plays a key role to produce magnetodielectric coupling. Dielectric constant measurements in presence of magnetic field can probe the coupling between charges and spins in insulating systems. Ising spin chain CoV$_2$O$_6$,\cite{ksingh} Ca$_3$Co$_2$O$_6$\cite{bell,bell1,basu} with triangular network display magnetodielectric coupling at low temperature. Quasi-one-dimensional Ising spin chain CoNb$_2$O$_6$ is a very good example of frustrated triangular lattice system which exhibits several interesting features like metamagnetic transition, quantum criticality behavior etc. Recently, quantum phase transition in transverse field has been experimentally evidenced in CoNb$_2$O$_6$.\cite{cold} A E$_8$ symmetry  has been experimentally observed near the quantum critical point of  Ising ferromagnet CoNb$_2$O$_6$. At low temperature, this system also exhibits various degenerate magnetically ordered states such as fourfold-degenerate antiferromagnetic (AF) phase, field-induced threefold-degenerate ferrimagnetic (FR) phase, sinusoidally amplitude-modulated incommensurate (IC) phase, confirmed by Neutron diffraction study.\cite{mit,koba,heid} In CoNb$_2$O$_6$ system, Co$^{2+}$ ions form zigzag chains along $c$-direction and they are arranged into isosceles triangular geometry in the $a-b$ plane. At low temperatures, Co spins orient along two different easy axes in the nearly $a-c$ plane with a 31$^{\circ}$ canting angle from the $c$ axis. Intrachain interaction is ferromagnetic in nature and chains are weakly coupled by antiferromagnetic interaction. In this paper, we have performed megnetodielctric measurement to evidence the coupling between electrical charges and magnetic moments. In addition, we have done magnetostriction measurements to probe coupling between spin and lattice. In some systems, spins are simultaneously coupled with both lattice and charges. For example, in EuTiO$_3$, magnetostriction measurement exhibits several similarities with the field dependence of the dielectric constant.\cite{reu} The correlation between spin-phonon coupling and dielectric constant has been observed in TbFe$_3$(BO$_3$)$_4$.\cite{adm} In this paper we have studied and compared magnetic, magnetodielectric, magnetothermal properties of CoNb$_2$O$_6$.
 \section{Experimental Details}
\begin{figure}[h!]
 	\begin{center}
 		\includegraphics[width=0.5\textwidth]{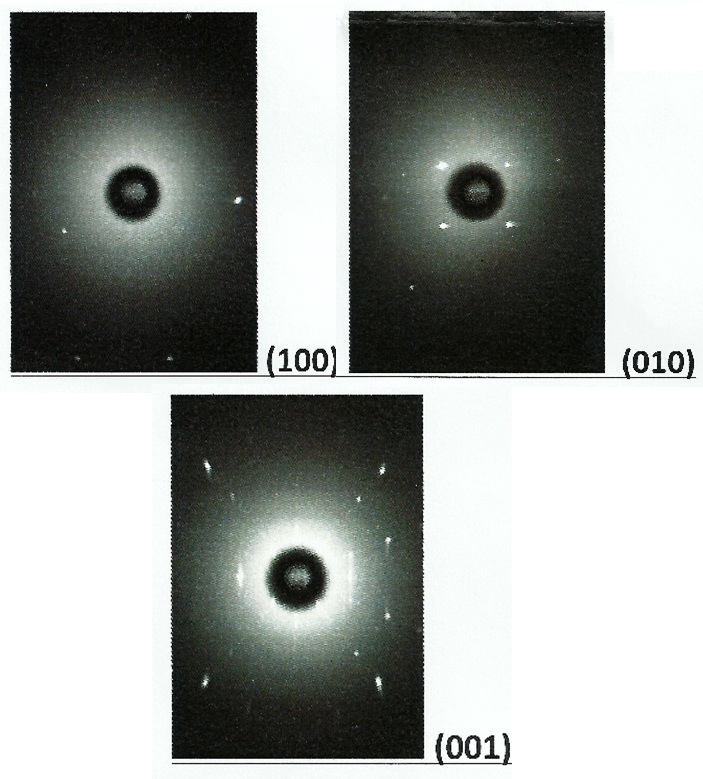}
 		\caption{Laue diffraction patterns of (100), (010) and (001) planes.}
 		\label{laue}
 	\end{center}
 \end{figure}

 Single crystal of CoNb$_{2}$O$_{6}$ was grown by the traveling solvent floating zone method.\cite{prabha} Laue XRD was performed to determine crystal axes and crystal was cut along different crystallographic planes according to experiment. Laue diffraction patterns are illustrated in Figure \ref{laue}. A rectangular shaped piece of single crystal was used for dielectric measurements. Two parallel faces of the crystal were covered with silver paint in order to apply an electric field perpendicular to the chains. Here electric field was applied along $a$ axis where as  magnetic field was applied along the easy axis direction $c$ so that the $\vec{E}$$\perp$$\vec{H}$ condition was always fulfilled. Magnetostriction measurements were done by capacitive method using a miniature tilted-plates dilatometer with applied field parallel to $c$ axis. The capacitance measurements were performed using a commercial AH2700A ultra-precision capacitance bridge. The magnetic measurements were done in SQUID-VSM (Quantum Design). The specific heat measurements were done using a physical property measurement system (Quantum Design) by conventional relaxation time method.
\section{Magnetization Measurements}
\begin{figure}[t!]
	\begin{center}
		\includegraphics[width=0.5\textwidth]{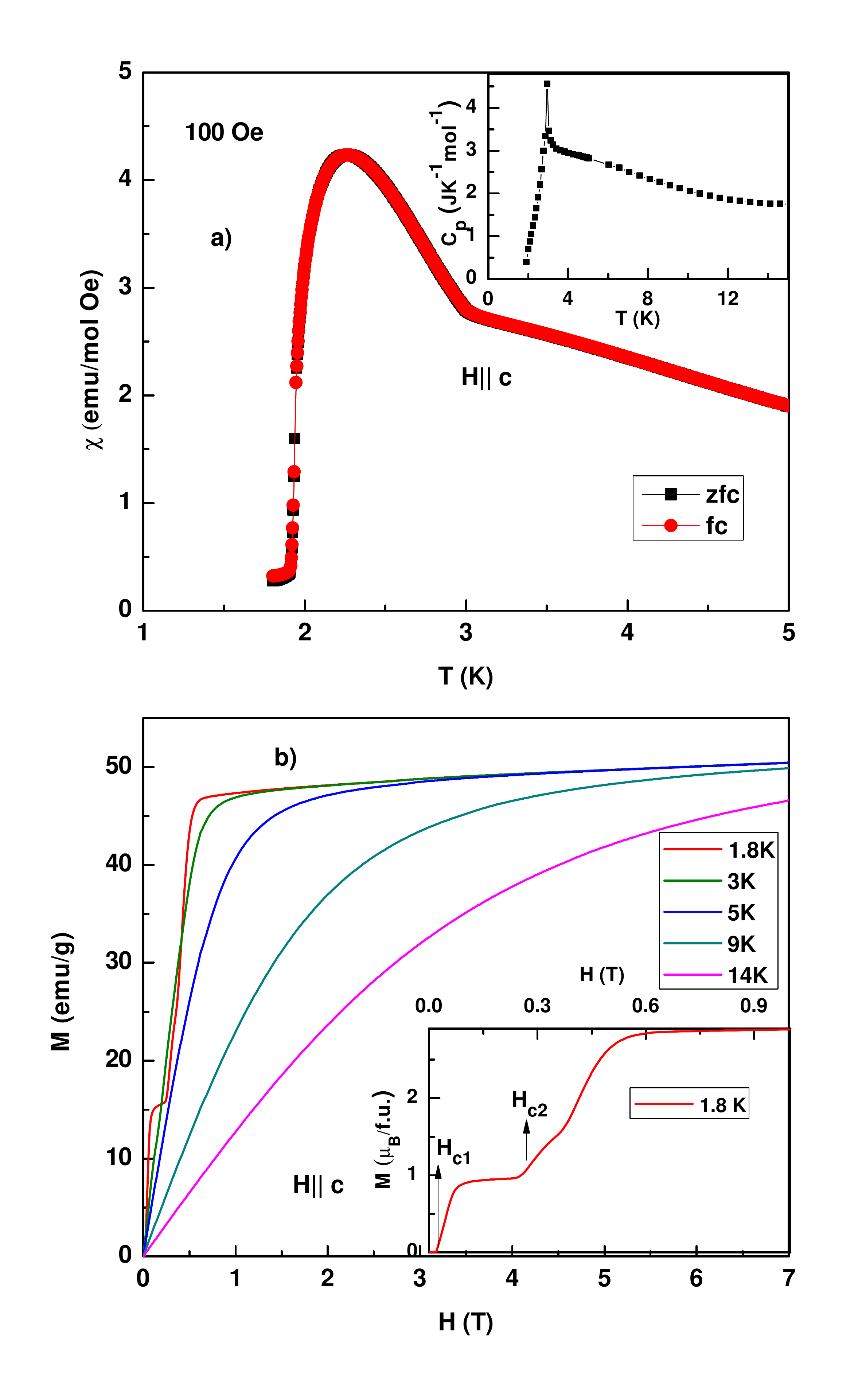}
		\caption{(a) Plots of $\chi$($T$) with zero field cool and field cool condition for CoNb$_2$O$_6$ at 100 Oe field applied along $c$ axis. Inset:Specific heat ($C_p$) versus temperature plot for CoNb$_2$O$_6$ at zero field. (b) Isothermal magnetization at some selected temperatures when field is applied along $c$ axis. Inset shows the closer view of M-H curve at 1.8 K.}
		\label{CN_M}
	\end{center}
\end{figure}

Temperature dependence of magnetization along $c$-axis in zero field cool(ZFC) and field cool(FC) conditions is plotted in Figure \ref{CN_M}(a). At low temperature, M vs. T curve shows two successive transitions below 3 and 2 K. ZFC and FC of M(T) do not show any bifurcation down to 1.8 K.  M(T) curve exhibits slope change around 3 K (T$_1$) due to transition from paramagnetic to incommensurate phase. Another transition occurs below 2 K (T$_2$) where a sharp drop in magnetization has been observed due transition from IC to AFM phase. In inset of Figure \ref{CN_M}(a), temperature dependence of specific heat ($C_p$(T)) is plotted at zero field. A very sharp peak has been observed at 2.9 K in $C_p$(T) due to PM-IC transition. Field dependence of magnetization along $c$-axis for some selected temperatures both above and below T$_1$ and T$_2$ are plotted in Figure \ref{CN_M}(b). In inset of Figure \ref{CN_M}(b), a closer view of M(H) at 1.8 K is given which exhibits multiple magnetization plateaux due to field induced magnetic phase transitions, similar to previously observed data.\cite{maar} M is very small up to 200 Oe then it shows step like jump at first critical field $H_{c1}$ and obtains 1/3 of saturation magnetization value in a certain field range. This step like increase in magnetization can be explained from magnetic phase diagram by S. Kobayashi.\cite{koba1} The saturation magnetization value is consistent with a Co$^{2+}$ moment of about 3 $\mu_B$. At 1.8 K, the system remains at AFM phase below 200 Oe, then it enters to ferrimagnetic phase via an incommensurate phase with increasing field. So this field induced AFM to ferrimagnetic phase transition is reflected in sharp step-like increase in M(H) curve around 200 Oe. Another increase in M(H) around 3.8 kOe (second critical field $H_{c2}$) is observed due to field induced transition from ferrimagnetic state to saturated PM state. With increasing temperature step-like increase in M(H) curve gradually disappears. Just above 3 K, magnetization linearly increases with H and then saturates. Further increase of temperature makes the M almost linear with H.

\section{Dielectric Constant Measurements}
\begin{figure}[h!]
	\begin{center}
		\includegraphics[width=0.50\textwidth]{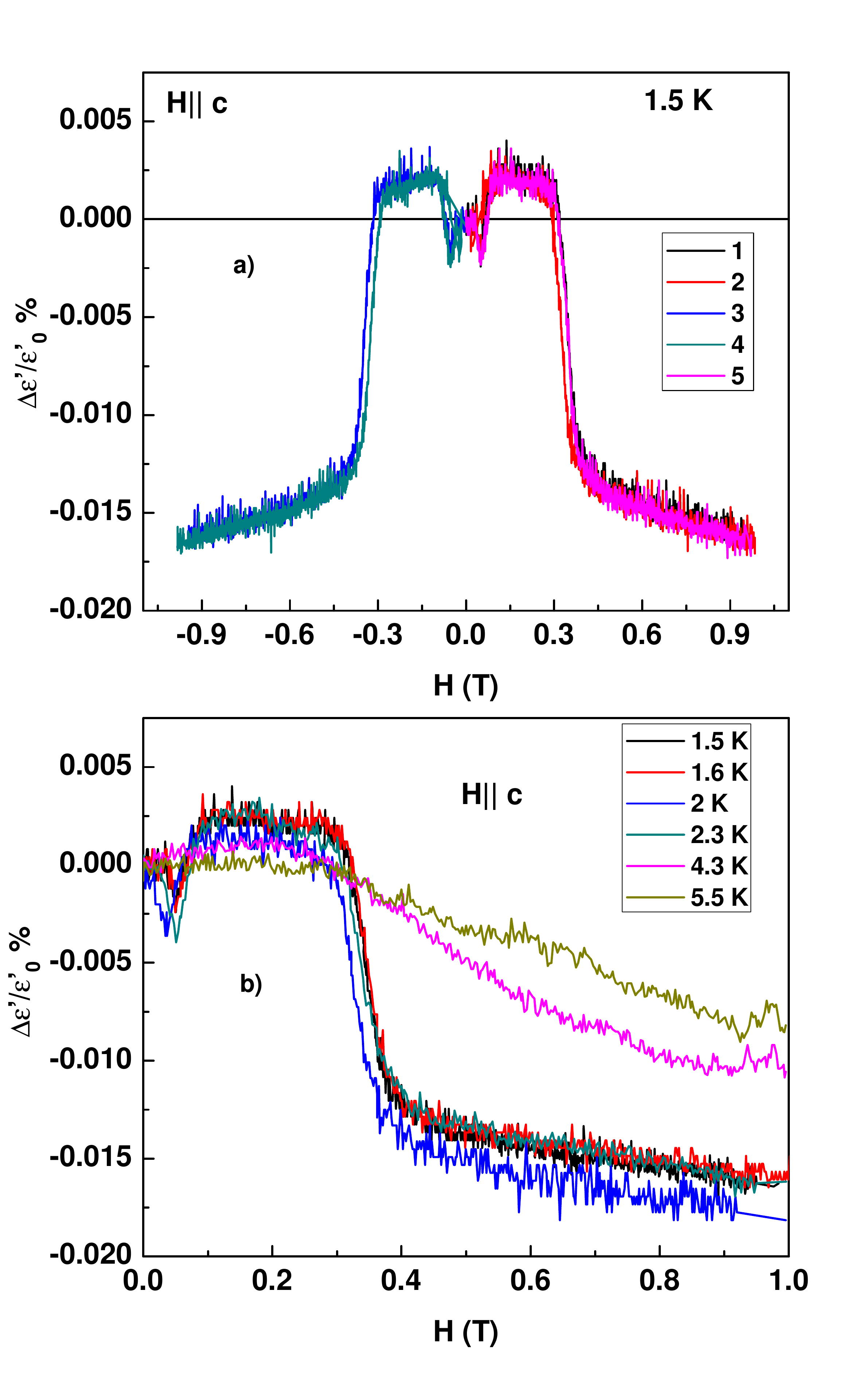}
		\caption{(a) Five segment curves for relative change of dielectric constant as a function of magnetic field are plotted at 1.5 K when magnetic  field applied along $c$ axis. (b) Plot of relative change of dielectric constant as a function of magnetic field for some selected temperatures when magnetic field is applied along $c$ axis.}
		\label{CN_di}
	\end{center}
\end{figure}
The isothermal dielectric constant measurements performed with a frequency 1 kHz, as a function of external magnetic-field at some selected temperatures are plotted in Figure \ref{CN_di}. Here external magnetic field is applied along $c$ axis and electric field is applied along $a$ axis. Actually, Figure \ref{CN_di} shows percentage of relative change in dielectric constant ($\Delta$$\varepsilon$$\prime$/$\varepsilon$$\prime$$_0$\ $=$[$\varepsilon$$\prime$($H$)-$\varepsilon$$\prime$$_0$]/$\varepsilon$$\prime$$_0$ ), where $\varepsilon$$\prime$$_0$  is the dielectric constant of the sample in absence of magnetic field. Sharp anomalies are found in isothermal relative change of dielectric constant at 1.5 K in five segment curve, shown in Figure \ref{CN_di}(a). With increasing field $\Delta$$\varepsilon$$\prime$/$\varepsilon$$\prime$$_0$\% remains constant initially and then exhibit a sharp negative peak around $H_{c1}$, then it shows almost a constant positive plateau region in a certain field range. With further application of magnetic field, $\Delta$$\varepsilon$$\prime$/$\varepsilon$$\prime$$_0$\% displays a sharp step-like jump around $H_{c2}$ and hysteresis has also been found here. Then, $\Delta$$\varepsilon$$\prime$/$\varepsilon$$\prime$$_0$\% decreases very slowly with increasing H. Depending on field strength, $\Delta$$\varepsilon$$\prime$/$\varepsilon$$\prime$$_0$\% obtains positive value as well as negative value. With increasing temperature, these anomalies gradually disappear, shown in Figure \ref{CN_di}(b). At 4.3 K, sharp peak-like feature around $H_{c1}$ is totally disappeared but $\Delta$$\varepsilon$$\prime$/$\varepsilon$$\prime$$_0$\% becomes positive in a certain field range though the plateau-like behavior is disappeared. No positive region has been found with further increase of temperature where dielectric constant monotonically decreases with increasing field. At 5.5 K, where system is above $T_1$, peak around $H_{c1}$ and step-like jump around $H_{c2}$ in dielectric constant disappear and it shows almost linear dependence with field. According to the magnetic phase diagram with field applied along $c$-axis, drawn from neutron diffraction data, CoNb$_{2}$O$_{6}$ undergoes multiple field induced magnetic transitions at 1.5 K.\cite{koba1} In low field region, below T$_2$, CoNb$_{2}$O$_{6}$  exhibits successive field induced antiferromagnetic (AF) to incommensurate (IC) then IC to ferrimagnetic (FR) transitions around 225 Oe and 395 Oe respectively. Dielectric constant shows two successive slope changes in low field region which can be interpreted by these AF to FR transition via intermediate IC phase. Above a certain critical field, this system enters to field induced FR state and remains in this state up to 3.2 T where corresponding dielectric constant exhibits a plateau-like feature obtaining a certain positive value. Observed sharp step-like jump in $\Delta$$\varepsilon$$\prime$/$\varepsilon$$\prime$$_0$\% around 0.33 T may be related to transition from field induced FR to saturated PM state. Field induced magnetic transitions are reflected in field dependent dielectric constant measurements also. So it is clear that dielectric constant is magnetically coupled in this system.

\section{Thermal Expansion Measurements}
\begin{figure}[h!]
	\begin{center}
		\includegraphics[width=0.5\textwidth]{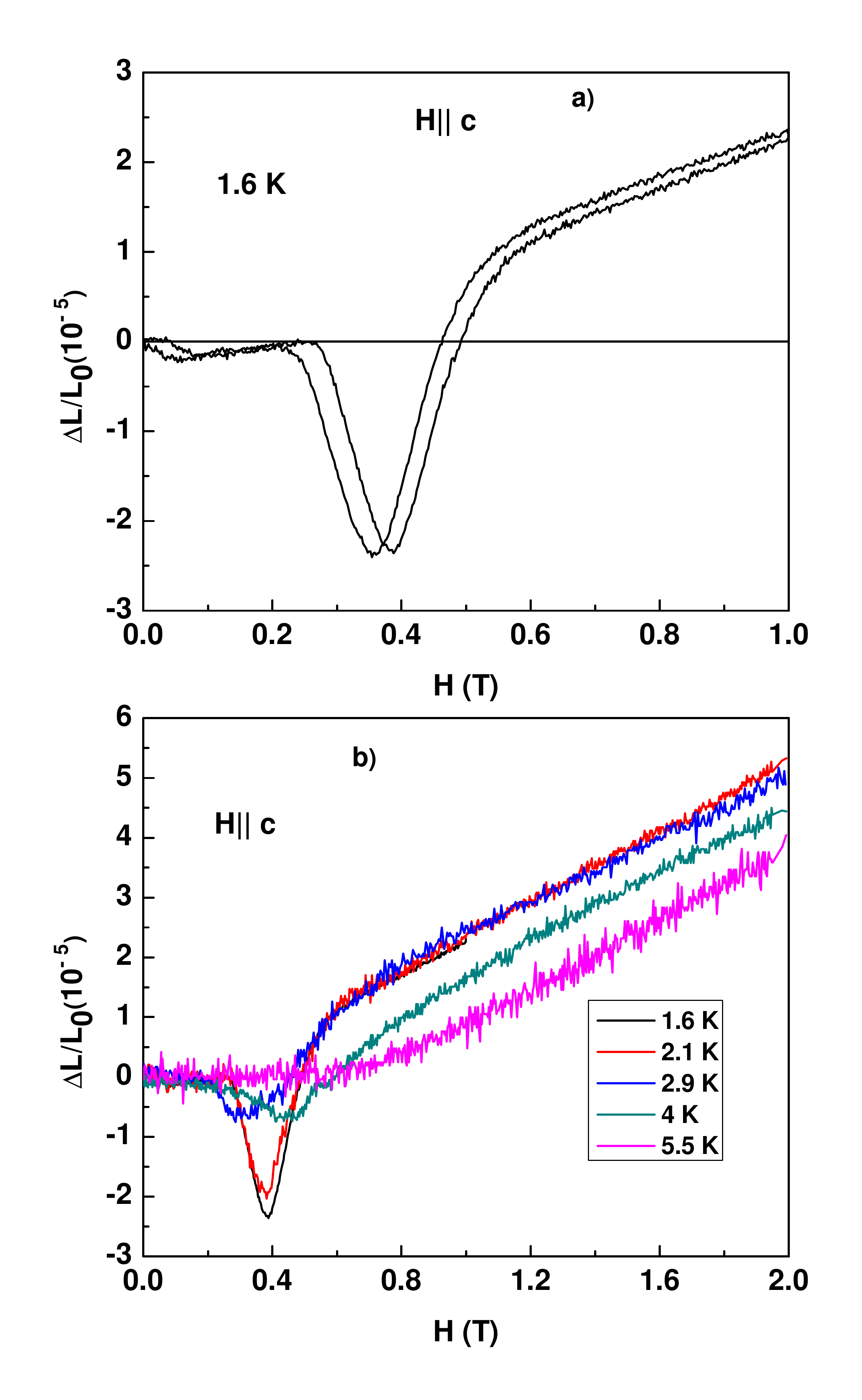}
		\caption{(a)Magnetostriction, $\Delta L(H)/L_0$ at 1.6 K for both field increasing and decreasing condition when magnetic  field applied along $c$ axis. (b) Magnetostriction, $\Delta L(H)/L_0$,  at several temperatures both above and below $T_1$ and $T_2$ for some selected temperatures when magnetic field is applied along $c$ axis.}
		\label{CN_th}
	\end{center}
\end{figure}
We have also performed magnetostriction measurements at some selected temperatures. Figure \ref{CN_th}(a) shows magnetostriction $\Delta L$($H$)/$L_0$$=$[$L$($H$)-$L_0$]/$L_0$, where  $L_0$ is the length of the sample  in absence of magnetic field, at 1.6 K for field increasing and decreasing conditions. Similar to dielectric constant measurement, magnetostriction at 1.6 K also exhibits two anomalies around $H_{c1}$ and $H_{c2}$. Hysteresis has also been found here. $\Delta L$($H$)/$L_0$ shows a weak cusplike anomaly around 500 Oe where a peak type feature has been found in dielectric constant measurement. A very pronounced peak has been found in $\Delta L$($H$)/$L_0$ around 3500 Oe where dielectric constant exhibits a very sharp step-like jump almost near about this field. The very sharp peak around 3500 Oe is associated with the transition from the ferrimagnetic state to a saturated PM high-field phase. Apart from this, magnetostriction measurement shows an interesting behavior. $\Delta L$($H$)/$L_0$ obtains positive value as well as negative value. Similar kind of behavior is also observed in magnetostriction of EuTiO$_3$ where it shows a sign change with increasing magnetic field\cite{reu}. With increasing field $\Delta L$($H$)/$L_0$ remains negative up to 4700 Oe then it becomes positive and increases linearly with field above 6000 Oe. Magnetostriction curves for some selective temperatures both below and above $T_1$ and $T_2$ are shown in Figure \ref{CN_th}(b).  At 2.1 K, weak anomaly around $H_{c1}$ disappears but sharp peak around $H_{c2}$ has been observed. With increasing temperature, height of the peak around 3500 Oe decreases gradually and disappears above 4 K. At 5.5 K, $\Delta L$($H$)/$L_0$ exhibits very small value up to 6000 Oe and then it increases monotonically with increasing field but it remains always positive throughout this region. Particularly at low temperature, field induced metamagnetic transitions are reflected in magnetostriction measurements also which suggests that spins are strongly coupled with lattice in this system.

\section{Summary}
We have carried out magnetostriction and magnetodielectric measurements on single crystalline CoNb$_{2}$O$_{6}$ sample at low temperature. The samples are well characterized by magnetization and specific heat measurements. We have related field dependence of dielectric constant and thermal expansion measurements with magnetic phase diagram by neutron diffraction and compared with magnetization. Multiple phase transitions observed by neutron diffraction data are clearly reflected in field dependence of dielectric constant and magnetostriction measurements. Field dependence of the dielectric constant display several similarities with magnetostriction measurements in CoNb$_{2}$O$_{6}$. Both field dependence of dielectric constant and magnetostriction exhibit two anomalies around two critical fields of metamagnetic transitions and obtain positive as well as negative value. So it can be concluded that spin-lattice coupling plays a key role and spins are magnetically coupled with charges in this system.

\end{document}